# Artificial Intelligence-based Smart Port Logistics Metaverse for Enhancing Productivity, Environment, and Safety in Port Logistics: A Case Study of Busan Port

Sunghyun Sim[a,†], Dohee Kim[b,†], Kikun Park[c,†], Hyerim Bae[d,‡]

[a]*Department of Industrial Management Big Data Engineering, Dongeui University,*
*Busanjin Gu, 47340 Busan, South Korea, ssh@deu.ac.kr*

[b]*Safe and Clean Supply Chain Research Center, Pusan National University,*
*Geumjeong-gu, Busan 46241, South Korea, kimdohee@pusan.ac.kr*

[c]*Department of Industrial Engineering, Pusan National University,*
*Geumjeong-gu, Busan 46241, South Korea, doublekpark@pusan.ac.kr*

[d]*Department of Data Science, Graduate School of Data Science, Pusan National University,*
*Geumjeong-gu, Busan 46241, South Korea, hrbae@pusan.ac.kr*

[†]*These authors contributed equally to this work,* [‡]*Corresponding author*

**Abstract.** The increase in global trade, the impact of COVID-19, and the tightening of environmental and safety regulations have brought significant changes to the maritime transportation market. To address these challenges, the port logistics sector is rapidly adopting advanced technologies such as big data, Internet of Things, and AI. However, despite these efforts, solving several issues related to productivity, environment, and safety in the port logistics sector requires collaboration among various stakeholders. In this study, we introduce an AI-based port logistics metaverse framework (PLMF) that facilitates communication, data sharing, and decision-making among diverse stakeholders in port logistics. The developed PLMF includes 11 AI-based metaverse content modules related to productivity, environment, and safety, enabling the monitoring, simulation, and decision making of real port logistics processes. Examples of these modules include the prediction of expected time of arrival, dynamic port operation planning, monitoring and prediction of ship fuel consumption and port equipment emissions, and detection and monitoring of hazardous ship routes and accidents between workers and port equipment. We conducted a case study using historical data from Busan Port to analyze the effectiveness of the PLMF. By predicting the expected arrival time of ships within the PLMF and optimizing port operations accordingly, we observed that the framework could generate additional direct revenue of approximately 7.3 million dollars annually, along with a 79% improvement in ship punctuality, resulting in certain environmental benefits for the port. These findings indicate that PLMF not only provides a platform for various stakeholders in port logistics to participate and collaborate but also significantly enhances the accuracy and sustainability of decision-making in port logistics through AI-based simulations.

## 1. Introduction

More than 80% of global trade by volume and more than 70% by value is conducted through port logistics (Verschuur, 2022). Over the past decade, global trade volume has consistently increased. Recent global issues, such as the COVID-19 pandemic, have accelerated the rise in trade volumes worldwide, resulting in a rapid increase in trade conducted by sea (Bai et al., 2022; Gu et al., 2023). As the volume of seaborne trade increases, the cargo flow within port logistics also increases. Particularly in ports that connect inland and maritime transportation, the involvement of various stakeholders, such as freight carriers, shipping companies, terminal operators, trucking companies, and government agencies, results in a more complex logistics network and reduces operational efficiency (Kuakoski et al, 2023). To address these issues, various information systems, such as port community systems (PCS) and port operating systems (POS), have been introduced to facilitate smooth information exchange among stakeholders and improve port operational efficiency (Muravev et al, 2021). Recently, the adoption of intelligent port information systems, such as those incorporating blockchain and AI, has contributed not only to reducing the complexity of logistics networks but also to enhancing operational efficiency.

Previously, improving the operational efficiency of port logistics was of primary importance. Environmental and safety issues have recently emerged as critical concerns in port logistics. First, there is a growing emphasis on carbon neutrality globally, resulting in stricter environmental regulations such as carbon trading and the IMO's emission control areas. Because these environmental regulations directly affect the profits of shipping companies, they must consider these factors in their decision-making. Additionally, changes, such as an increase in ship size and cargo volume, can impact safety. Large vessels pose greater risks owing to potentially more catastrophic accidents, which can result in severe economic losses, environmental damage, and potential casualties. Increased cargo volumes have been observed to result in higher workloads and corresponding increases in accident frequency. An increase in cargo volume also contributes to a higher workload at ports, resulting in a greater frequency of accidents due to human error. Owing to these changes, there is increasing emphasis on the need to develop port logistics information systems that can address environmental and safety-related issues. (Liu et al., 2019; Raza, 2020; Khan et al., 2023; Wu et al., 2022).

Recent research on addressing environmental and safety issues within global supply chains indicates that the metaverse could overcome the limitations of existing information systems. Sarwatt et al. (2023) presented a case study on the adoption of the metaverse to address environmental and safety issues in general logistics environments. Specifically, they proposed a model that integrates autonomous vehicles within logistics environments to optimize decision-making, provide optimal routes for vehicles, and reduce energy consumption, thereby mitigating environmental impact.

Table 1. Summary of studies related to the adoption of the metaverse in the port logistics area.

| Research | Research Type | Contents | | | Coverage | | | Additional Tech |
| --- | --- | --- | --- | --- | --- | --- | --- | --- |
| | | *Efficient Issue* | *Environments Issue* | *Safety Issue* | *Sea* | *Sea-Port* | *Port* | |
| Cabrero et al. (2024) | Conceptual | ∨ | ∨ | ∨ | ∨ | ∨ | ∨ | Blockchain, Optimization |
| Nicoletta et al. (2024) | Conceptual | ∨ | ∨ | | | | | AI |
| Mário et al. (2023) | Conceptual | ∨ | ∨ | ∨ | ∨ | | | AR/VR |
| Alexandre and Ivanov (2023) | Conceptual | ∨ | | ∨ | | | ∨ | - |
| Yang et al. (2023) | Conceptual | ∨ | | | ∨ | | | Blockchain, AI, VR |
| Deveci et al. (2022) | Conceptual | ∨ | ∨ | | | | ∨ | Optimization |
| **Ours** | **Empirical** | ∨ | ∨ | ∨ | ∨ | ∨ | ∨ | **AI** |

In the field of port and port logistics, studies related to the adoption of the metaverse include those by Alexandre and Ivanov (2023), Mário et al. et al. (2024), and Deveci et al. (2022). Alexandre and Ivanov (2023) proposed the adoption of a metaverse approach to address safety issues in port logistics environments. They demonstrated that the metaverse model can be used to provide emergency response training and virtual simulations for risk assessment, which can address safety issues in port logistics. In particular, they showed that real-time virtual safety audits and risk monitoring can improve safety measures. Mário et al. (2024) discussed a case study on the adoption of a metaverse strategy to enhance safety in ports and port logistics. Their research demonstrated that immersive training simulations for workers, real-time monitoring for safety management, and virtual scenario planning for emergency response drills could significantly improve safety practices. Additionally, they proposed an architecture that utilizes advanced technologies, such as extended reality and artificial intelligence, within the metaverse to enhance decision-making processes, optimize resource utilization, and mitigate risks in port operations. Deveci et al. (2022) focused on an integrated assessment of the metaverse to measure cargo flow. Although their study did not directly address environmental issues in the maritime sector, they demonstrated that using a metaverse for cargo measurement could potentially result in more efficient logistics operations, ultimately reducing environmental impacts through route optimization, fuel consumption, and carbon emissions. In addition, as summarized in Table 1, several studies address the adoption of the metaverse in port logistics; however, most remain at the conceptual level, which is a limitation.

This study introduces the port logistics metaverse framework (PLMF), which collects, preprocesses, and models real data to practically implement the entire process occurring in port logistics. This framework simultaneously addressed efficiency, environmental, and safety issues. In addition, the PLMF incorporates AI technology to propose an approach that enhances the accuracy of various decision-making processes in port logistics operations.

The main contributions of this study are as follows:

- We developed a metaverse framework capable of decision-making across the entire process in maritime and port logistics sectors.

- Within the developed metaverse framework, various environmental and safety issues in the maritime and port logistics sectors can be monitored in real time, and the accuracy of the decision-making process can be enhanced through an AI-based decision-making system.

- Using PLMF, we predicted the arrival times of vessels entering and leaving Busan Port and conducted simulations of port logistics processes based on these predictions. The results demonstrated a significant increase in annual direct revenue and substantial improvements in on-time ship arrivals, leading to environmental benefits for the port.

The remainder of this paper is organized as follows: The structure of the proposed metaverse framework is presented in Section 3. Section 4 presents scenarios and functionalities within the developed metaverse from the perspectives of productivity, environment, and safety. Section 5 outlines the results of a case study conducted at the Busan Port using the developed metaverse. Finally, Section 6 consolidates the findings of this study and explores potential directions for future research.

## 2. AI-based port logistic metaverse framework

In this section, we introduce the AI-based PLMF. Figure 1 illustrates the overall structure of the PLMF. The proposed framework considers five key processes that occur in real port logistics: 1) Departure: The process of finishing operations at the previous port and departing; 2) Ship Operation: The process of the vessel navigating towards its destination; 3) Berthing: The process of the vessel arriving at its destination and docking at the berth using tugboats; and 4) Unloading and Loading: The transferring cargo from the ship to yard trucks or from yard trucks to the ship. 5) Moving: The process of moving the cargo stored in the yard to the ship or transporting unloaded cargo from the ship to the yard.

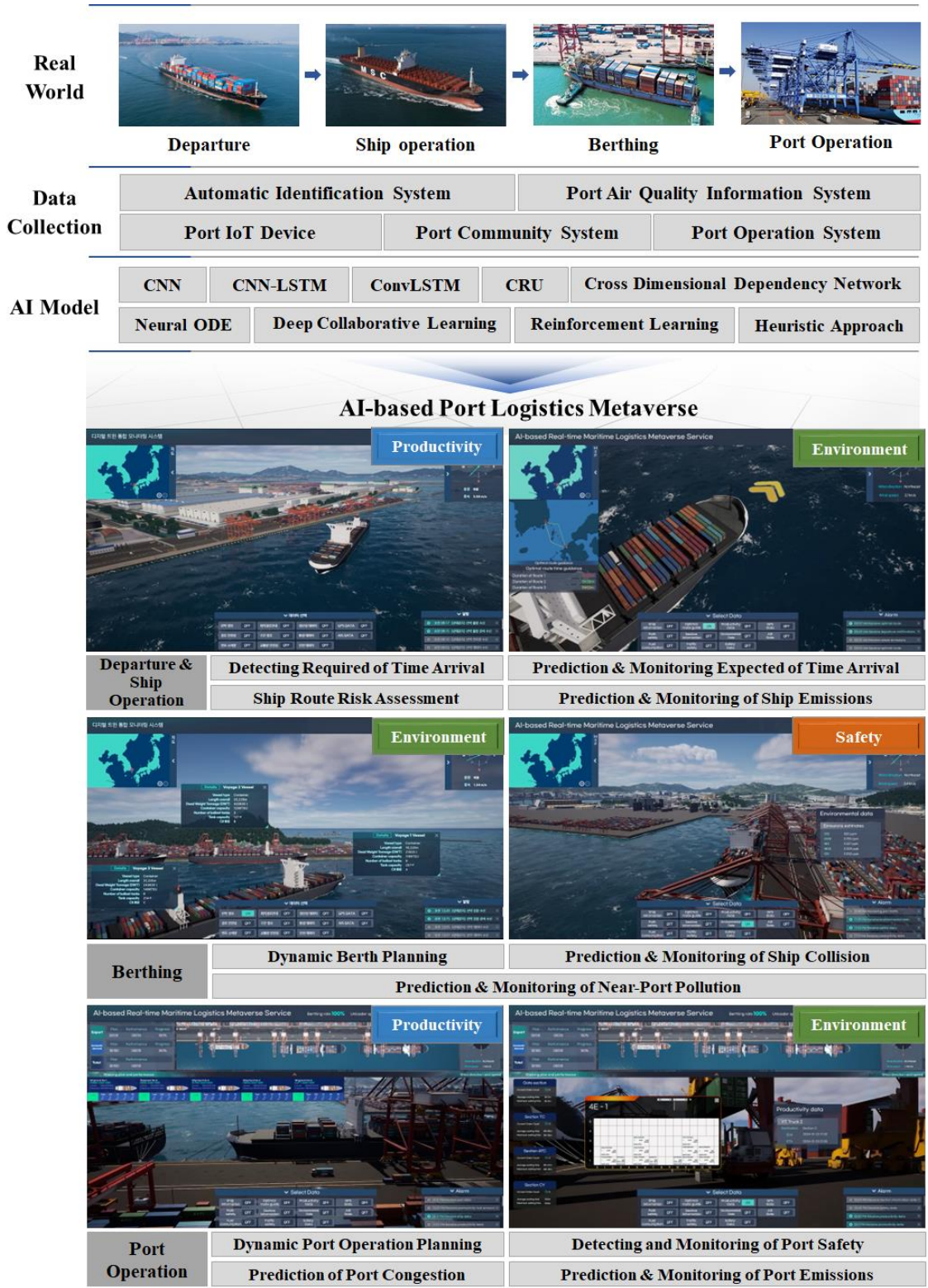

Figure 1. Overall framework of AI-based maritime-port logistic metaverse.

The PLMF collects structured and unstructured data from the systems used in each process. The collected data are subsequently integrated and preprocessed according to their specific functions. After the preprocessing stage, the resulting data are used as follows. 1) Some of the data are used for monitoring. (2) Other data are used for prediction and detection using AI models. Ultimately, within the metaverse, the real-time data collected from the five processes are monitored, and various prediction and detection insights derived from the AI models are visualized

In Section 2.1, we introduce the systems used for data collection and describe the attributes of the data collected from these systems. In Section 2.2, we present the detailed functionalities provided by the PLMF for each process scenario (monitoring, prediction, and detection) and introduce the data and AI models used in each scenario.

## 2.1. Data collection

In this section, we introduce the data used in PLMF. Table 2 summarizes the objects, key systems, and data attributes collected from each system during the five processes. These systems include those operated by shipping companies, vessels, tugboats, quay cranes, yard trucks, yard cranes, port managers, workers, and authorities. The systems from which data are sourced include the automatic identification system (AIS), port Internet of Things (IoT) devices, the port air quality monitoring system (PAQMS), PCS, POS, and CCTV.

Table 2. Summary of the related process, object, system, and data attribute for developing AI-based PLMF.

| Location | Real World Process | Object | System | Data Attribute |
|---|---|---|---|---|
| Sea | Departure Ship Operation | • Ship | • Automatic Identification System | • Maritime Mobile Service Identity<br>• Timestamp<br>• Latitude<br>• Longitude<br>• Speed Over Ground<br>• Course Over Ground<br>• Heading<br>• Rate of Turn<br>• Ship Type<br>• Ship Length<br>• Ship Width |
| Sea & Port | Berthing | • Ship<br>• Pilot Ship | | |
| Port | Port Operation (Container Unloading & Loading + Moving) | • Quay Crane<br>• Yard Truck<br>• Yard Crane<br>• Worker | • IoT Device | • Equipment Index<br>• Device Identify<br>• Timestamp<br>• Latitude<br>• Longitude<br>• Altitude<br>• Velocity<br>• Direction<br>• Work Type |
| | | | • Port Air Quality Monitoring System | • Wind Direction<br>• Wind Speed<br>• Humidity<br>• Air Quality Index ($PM_{2.5}$, $PM_{10}$, $NO$, $NO_x$, $SO$, $SO_2$, |

|   |   |   |
|---|---|---|
|   |   | $CO$, $CO_2$, $O_3$) |
|   | • Port Community System | • Vessel Traffic Service Status<br>• Terminal/Facility Utilization Status<br>• Dangerous Goods Handling Status |
|   | • Port Operation System | • Container Identify<br>• Vessel Identify<br>• Voyage Number<br>• Berth Number<br>• Vessel Arrival / Departure Time<br>• Container Type/Size<br>• Yard Location<br>• Job Type<br>• Gate In/Out Time<br>• Resource Identify<br>• Hazardous Material Indicator<br>• Service Type |
|   | • Port CCTV | • CCTV Video |

### 2.1.1. Data collected from the AIS

The AIS contains important information for navigation safety and maritime security and is an international standard for information and communication between ships (Tetreault, 2005). The AIS is an essential device installed on 99% of ships worldwide. In the port logistics domain, all aspects related to vessels utilize data collected from the AIS.

| Timestamp | MMSI | Latitude | Longitude | SOG | COG | Heading | ROT | Draught | Ship Type | Ship Length | Ship Width |
|---|---|---|---|---|---|---|---|---|---|---|---|
| 2019-07-03 00:00:15.015121 UTC | Ship$_1$ | 35.09359667 | 129.0357483 | 0 | 0 | 137 | 0 | 4.4 | 52 | 30 | 10 |
| 2019-07-03 00:00:16.642674 UTC | Ship$_2$ | 35.08949667 | 129.0261833 | 0 | 0 | 511 | -128 | 3.5 | 52 | 40 | 9 |
| 2019-07-03 00:00:19.853931 UTC | Ship$_3$ | 35.09328000 | 129.1129650 | 8.8 | 226.4 | 511 | -128 | 4 | 89 | 84 | 13 |
| 2019-07-03 00:00:22.569505 UTC | Ship$_4$ | 35.09366333 | 129.0315883 | 0 | 0 | 511 | -128 | 7.3 | 73 | 144 | 22 |
| 2019-07-03 00:00:24.191044 UTC | Ship$_5$ | 35.11697167 | 129.0509133 | 0 | 341.4 | 308 | 0 | 4 | 80 | 62 | 11 |
| 2019-07-03 00:00:24.910697 UTC | Ship$_6$ | 35.10366000 | 129.0922233 | 6 | 276.5 | 511 | -128 | 3.5 | 30 | 53 | 9 |
| 2019-07-03 00:00:28.455377 UTC | Ship$_7$ | 35.09622000 | 129.0399350 | 0 | 276.1 | 511 | -128 | 6.1 | 80 | 101 | 17 |
| 2019-07-03 00:00:31.130697 UTC | Ship$_8$ | 35.47698167 | 129.3860750 | 0.1 | 238.8 | 254 | -128 | 5 | 80 | 40 | 9 |
| 2019-07-03 00:00:32.547593 UTC | Ship$_9$ | 35.49760333 | 129.3865383 | 8.8 | 121.4 | 511 | -128 | 3.6 | 30 | 39 | 8 |
| 2019-07-03 00:00:41.189244 UTC | Ship$_{10}$ | 35.07255000 | 129.1498500 | 12.2 | 278 | 271 | 0 | 4 | 52 | 33 | 9 |

⋮

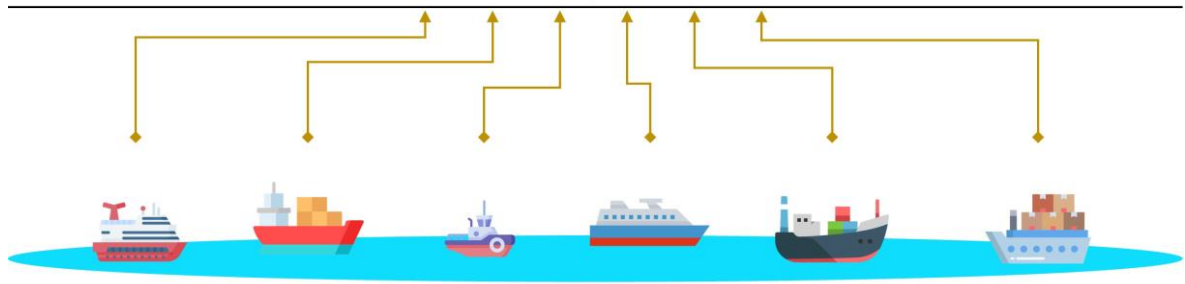

Figure 2. Example of data collected from an AIS.

Data collected from the AIS include static and dynamic information (Park et al., 2021). The static information included maritime mobile service identity and vessel specifications (Ship Type, Ship Length, Ship Width). Dynamic information includes information that can estimate the movement path and location of a ship, such as location information (latitude and longitude), speed over ground, course over ground, heading, and rate of turning (Sim et al., 2022). Figure 2 shows a sample of the AIS data collected from several ships. These data were used to monitor and predict various details related to vessels in the metaverse, such as the location of the vessels, emission levels, and safety levels during the departure/ship operation/berthing processes.

**2.1.2. Data collected from port IoT devices**

The IoT is a rapidly evolving global networking infrastructure in modern wireless communication that enables the collection, management, processing, and distribution of data through devices and physical objects by interconnecting different networks (Paul et al, 2020; Alex et al,.2022; Reza et al., 2024). The introduction of IoT at ports offers benefits such as improved efficiency, safety, and port automation, leading to many projects aimed at implementing IoT (Yang et al., 2018). Consequently, many major global ports, such as Busan, Hamburg, Rotterdam, and Long Beach have equipped most of their equipment with IoT technology, enabling real-time data collection (Yau et al., 2020; Choi et al., 2021; Kapkaeva et al., 2021; Gaspare et al., 2021).

| Timestamp | Equipment Index | Device Identify | Latitude | Longitude | Altitude | Velocity | Direction | Work Type |
|---|---|---|---|---|---|---|---|---|
| 2021-10-31 T20:59:59Z | YT$_1$ | 1 | 35.1078 | 129.0972 | 14 | 2 | 193 | U |
| 2021-10-31 T21:00:01Z | YT$_2$ | 2 | 35.1020 | 129.0963 | 11 | 18 | 183 | L |
| 2021-10-31 T21:00:08Z | YT$_3$ | 3 | 35.1016 | 129.0963 | 11 | 17 | 174 | U |
| 2021-10-31 T20:59:59Z | QC$_1$ | 4 | 35.1078 | 129.0972 | 14 | 0 | 0 | U |
| 2021-10-31 T21:00:14Z | QC$_2$ | 5 | 35.1078 | 129.0973 | 14 | 0 | 0 | U |
| 2021-10-31 T21:00:31Z | QC$_3$ | 6 | 35.1079 | 129.0974 | 14 | 0 | 0 | L |
| 2021-10-31 T21:00:01Z | YC$_1$ | 7 | 35.1020 | 129.0963 | 11 | 0 | 0 | L |
| 2021-10-31 T21:00:06Z | YC$_1$ | 8 | 35.1010 | 129.0969 | 11 | 0 | 0 | U |
| 2021-10-31 T21:00:09Z | Work$_1$ | 9 | 35.1025 | 129.0965 | 11 | 3 | 185 | U |
| 2021-10-31 T21:01:12Z | Work$_2$ | 10 | 35.1043 | 129.0968 | 11 | 1 | 185 | L |

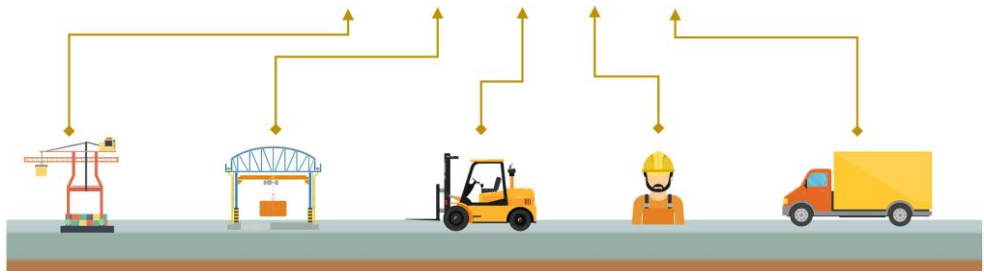

Figure 3. Example of data collected from an IoT device.

Figure 3 shows examples of data collected from IoT devices installed on equipment such as quay cranes (QC), yard cranes (YC), and yard trucks (YT) at Busan Port. Each data attribute includes the equipment index, identifier of the IoT device attached to each piece of equipment, location information (latitude, longitude, altitude, velocity, and direction), and type of operation (U: container unloading job, L: container loading job) of the equipment. These data are used to monitor, predict, and schedule various details related to port operations in the metaverse during berthing/loading and unloading processes. For instance, they are directly utilized for operational planning (berth allocation, QC allocation and scheduling, YC scheduling, etc.), monitoring and predicting emission levels at ports, and monitoring and predicting safety levels.

### 2.1.3. Data collected from PAQMS

Port areas are known for being spaces where various equipment, including ships, operate, and are significant sources of pollution (Wan et al., 2022). Pollution levels from port logistics account for approximately 3% of global greenhouse gas emissions, with nitrogen oxides ($NO_x$) and sulfur oxides ($SO_x$) contributing 13 and 12%, respectively (Mueller et al., 2023). Additionally, harmful pollutants such as particulate matter (PM), black carbon (BC), and methane ($CH_4$) are also emitted, posing risks to human health. Therefore, ports have implemented systems to continuously monitor and manage pollutant emissions (Mocerino et al., 2020; María et al., 2024).

PAQMSs collect data on pollutants such as PM, fine PM ($PM_{2.5}$), NO, $NO_2$ $NO_x$, SO, $SO_2$, CO, $CO_2$, and $O_3$ as well as meteorological information, including temperature, humidity, wind direction, and wind speed. These data are used to monitor and predict pollutant emissions from metaverse ports.

### 2.1.4. Data collected from PCS and POS

A PCS is an information system that facilitates networking between public and private organizations and entities involved in providing vessel and cargo services at the port (Iida and Daisuke, 2023). Although they vary by country, there are generally three types of PCS: 1) A one-stop service system "*that covers maritime and port administrative procedures, such as port entry/departure declaration, notice of security reports, and other related information between private sectors and public authorities nationwide*" (IMO, 2021). Countries that use these types of PCS include South Korea (Port-MIS), Singapore (DigitalPort@SG), Sweden (Reportal), Spain (DUEPORT), Germany (NSW Deutschland), the Netherlands (SWM&A), and Japan (NACCS); 2) A one-stop service system "*that covers procedures related to exports and imports goods such as customs clearance*" (IMO, 2021). Countries that use these types of PCS include United States (ACE), United Kingdom (CDS), and Singapore (TradeNet); 3) A one-stop service system "*that electronically processes document exchanges, work orders, or sharing visibility data on commercial services for the operation of maritime and port logistics between stakeholders within a port community*" (Moros et al., 2020). Countries that use these types of PCS include Netherlands (Portbase), Germany (DAKOSY), Singapore (PORTNET), South Korea (PLISM), and United Kingdom (Destin8). Data derived from the PCS generally include control and specification information of arriving and departing vessels, cargo status information, and details of the usage of port facilities.

A POS (or terminal operating system, TOS) is an information system designed to manage the operations and logistics within a port. Its main functions include managing the movement, storage, and tracking of containers and other cargo within the terminal, optimizing yard operations, and coordinating the use of equipment, such as cranes and trucks (Hus et al., 2023). The data attributes collected by the POS include container identity, vessel identity, voyage number, berth number, vessel arrival/departure time, container type/size, yard location, job type, gate in/out time, resource identity, hazardous material indicator, and service type.

These data were used within the metaverse to monitor and optimize the scheduling of operations (such as loading and unloading) within the port. This includes scheduling from the perspective of facilities, equipment, and tasks, and data are utilized to simulate these processes to improve efficiency.

## 2.2. PLMF contents with AI

Within the PLMF, not only are various productivity, environmental, and safety issues arising in port logistics processes monitored but AI models are also used to optimize, predict, and detect related issues. This section introduces the specific problems addressed by the PLMF, and the data and models used to solve these problems. Table 3 summarizes the key content addressed by the PLMF and the stakeholders related to each content. Table 4 summarizes the effects of each content on issues related to productivity, environment, and safety. As listed in Table 3, PLMF not only monitors the entire port logistics process but also involves predicting, detecting, and real-time optimization of 12 different aspects related to productivity, environment, and safety at each procedure. These processes can positively affect the stakeholders involved in port logistics. Furthermore, they offer advantages in terms of enhancing productivity and addressing environmental and safety issues. In Sections 2.2.1–2.2.3, we will introduce the specific problems that PLMF aims to solve for each detailed content and the data and models used to address these problems.

Table 3. Summary of PLMF contents, effectivity, and related business stakeholders.

| Process | Contents | Effectivity | | | Related Business Stakeholder |
|---|---|---|---|---|---|
| | | Productivity Effect | Environment Effect | Safety effect | |
| Departure & Ship Operation | Detecting Require of Time Arrival | ∨ | | | Shipping Company Shipper Port Operator |
| | Prediction Expected of Time Arrival | ∨ | | | |
| | Prediction & Monitoring of Ship Emissions | | ∨ | | |
| | Ship Route Risk Assessment | | | ∨ | |
| Berthing | Prediction & Monitoring of Ship Collision | | | ∨ | Shipping Company Marine Pilot Port Operator Trucking Company Port Authority |
| | Dynamic Berth Planning | ∨ | | | |
| | Prediction & Monitoring of Near-Port Pollution | | ∨ | | |
| Port Operation | Dynamic Port Operation Planning | ∨ | | | Shipping Company Shipper Port Operator Trucking Company Port Authority |
| | Prediction of Port Congestion | ∨ | | | |
| | Prediction & Monitoring of Port Emissions | | ∨ | | |
| | Detecting and Monitoring of Port Safety | | | ∨ | |

### 2.2.1. Departure, ship operation, and berthing

In the process of ship departure, ship operation, and berthing, the following seven issues are addressed. 1) *Detection of Arrival Time Requirements*: This issue aims to determine whether a ship can adhere to its originally estimated arrival time (ETA) based on its current position and current time. 2) *Prediction* *of Estimated Arrival Time*: This issue arises when the ship cannot meet the original ETA, necessitating the recalculation of the arrival time. Even if a new ETA is predicted, a ship may need to wait upon arrival if no berths are available at the port. Therefore, a new ETA is predicted by considering an existing berth schedule. 3) *Prediction of Ship Emissions*: This study aims to predict the emissions at the time of ship operation and the total emissions along the route. In this case, emissions may vary depending on the speed and route of the ship. 4) *Ship Route Risk Assessment*: This study aims to evaluate the risks associated with the route used by the ship, such as collisions or grounding in the area where the ship is currently operating. This assessment can be performed for either the entire route of the ship or for localized risks in the current operational area. 5) *Prediction of Ship Collision*: To prevent collisions, this study aims to predict the risk of collisions between ships in congested coastal areas based on time and space. In this case, the ship domain, which represents the safe distance between ships (considering direction and speed), was generated from the AIS data. The generated ship domain is subsequently used to predict the collision risk between ships. 6) *Prediction of Near-Port Pollution*: This study aims to predict the level of pollution in coastal areas, anchorages, and berths. When a ship arrives at a port but cannot dock at a berth, it waits at an anchorage where it continues to emit pollutants. Pollution is also emitted from berth cranes and yard trucks positioned in the berths. This information was used to predict the pollution levels near the port. 7) *Dynamic Berth Planning*: This study aims to allocate berths to ships dynamically as their ETA changes in real time.

Table 5. Summary of key contents, utilized data, and models used in ship departure, operation, and berthing processes.

| Contents | Data used | Models used in PLMF |
| --- | --- | --- |
| Detecting Require of Time Arrival | AIS Data<br>Expected of Time Arrival | CNN-LSTM (Abdi and Amrit, 2024)<br>ConvLSTM<br>Cross-Dimensional Dependency Network[†](Sim et al., 2024)<br>Heuristic Approach (Kwun and Bae, 2021)<br>Reinforcements Learning (Park et al., 2021) |
| Prediction Expected of Time Arrival | AIS Data<br>Berthing Planning | |
| Prediction of Ship Emissions | AIS Data<br>Ship Route<br>Ship Emission | |
| Ship Route Risk Assessment | AIS Data<br>Ship Route | |
| Prediction of Ship Collision | AIS Data | Ship Domain + CNN |
| Prediction of Near-Port Pollution | AIS Data<br>Port Air Quality System<br>Port Community System<br>Port Operating System | Deep Collaborative Learning (Sim et al., 2022) |
| Dynamic Berth Planning | AIS Data<br>Port Community System<br>Port Operating System | Heuristic Approach<br>Reinforcements Learning |

Table 5 summarizes the data and models used to address the seven aforementioned issues. For all seven issues, the AIS data were the most important. Additionally, ETA information, berth planning, ship routes, PCS, and POS were used.

### 2.2.2. Port Operation

The following four issues are addressed in the port operation process: 1) *Dynamic Port Operation Planning*: This focuses on developing real-time plans necessary for port operations, including container loading and unloading schedules and equipment allocation within the port. 2) *Prediction of Port Congestion*: This aims to forecast waiting times that occur during container handling at ports. Generally, as the number of containers handled with limited resources increases, waiting times also increase, significantly reducing port productivity. 3) *Prediction of Port Emissions*: This study estimates the total emissions produced within ports. This involved predicting the pollutants emitted by various equipment in the port based on the obtained information. 4) *Detection of Port Safety*: This focuses on identifying potential safety problems that may arise during cargo-handling operations at ports. It involves detecting issues such as collisions between equipment, equipment and personnel, cargo and equipment, and cargo and personnel, before they occur.

Table 6. Summary of key contents, data, and models used in port operation processes.

| Contents | Data used | Models used in PLMF |
|---|---|---|
| Dynamic Port Operation Planning | IoT Device<br>Port Community System<br>Port Operating System | Heuristic Approach (Hanif et al., 2023)<br>Reinforcements Learning† (Adi et al., 2021) |
| Prediction of Port Congestion | AIS Data<br>IoT Device<br>Port Operating System | Correlation Recurrent Units†(Sim et al., 2023)<br>Discrete Event Simulation† (Park et al. 2024)<br>Time-series Decomposition and Two-stage Attention† (Kim et al., 2022) |
| Prediction of Port Emissions | IoT Device<br>Port Air Quality System | Deep Collaborative Learning (Sim et al., 2022) |
| Detection of Port Safety | IoT Device<br>CCTV | Neural Ordinary Differential Equations<br>YOLO (Xu et al., 2022) |

Table 6 summarizes the data and models used to address these four issues. For all seven issues, IoT devices data are the most important. Additionally, PCS, POS, AIS data, PAQMS, and CCTV data were used. Models marked with a † in the table are AI models developed in-house, while models not marked were implemented based on other research findings.

## 3. Major development results obtained in PLMF service scenarios

This section introduces a metaverse service designed to address environmental and safety issues in port logistics. The prposed metaverse service integrates AI models to monitor, detect, and predict crucial information reported in the industry. The key features of the metaverse service include the productivity perspective (***prediction of estimated times of arrival*** and ***monitoring and optimization of port operations***), environment perspective (***monitoring and prediction of ship fuel consumption*** and ***monitoring of port equipment emissions***), and safety perspective (***risk assessment of ship routes***, ***detection and monitoring of dangerous ship routes*** and ***detection and monitoring of safety accidents involving workers and equipment***)

### 3.1. PLMF service scenarios from a productivity perspective

### 3.1.1. Prediction of ETA

From the perspective of port logistics and operations, the ETA of a vessel is a critical decision variable for determining the berth time when a vessel arrives at a port. Additionally, accurate ETA prediction enables inference of the types and durations of loading and unloading operations, such as cargo handling and stowage activities. Therefore, accurate ETA prediction is crucial from a productivity perspective. Typically, owing to the minimal speed variations during navigation, the ETAs of ships can vary significantly based on the meteorological and maritime conditions along their route. Hence, it is important to accurately predict weather conditions in maritime zones traversed by vessels. In the proposed metaverse service, we utilized spatiotemporal AI methods, such as convolutional long short-term memory (ConvLSTM), to accurately predict the conditions of maritime zones (Bi, Jinqiang, et al., 2024). By leveraging this information, we offer a service to predict the ETA, aiding in the efficient planning and management of maritime operations.

Figure 1 illustrates the process of predicting the ETA of a ship within a metaverse to provide information to metaverse users. The primary users of this system are ship navigators and berth-operation schedulers. The movement direction of the ship and its heading are shown at the center of the figure. In the top-left corner, the current location information of the ship is displayed, with the optimal routes to the destination shown below along with the remaining time for each route. The top-right corner provides weather information, and the bottom section provides notifications of the necessary information. This system not only informs users of the predicted ETA but also interacts with them to calculate the optimized ETA in real time based on the selected route.

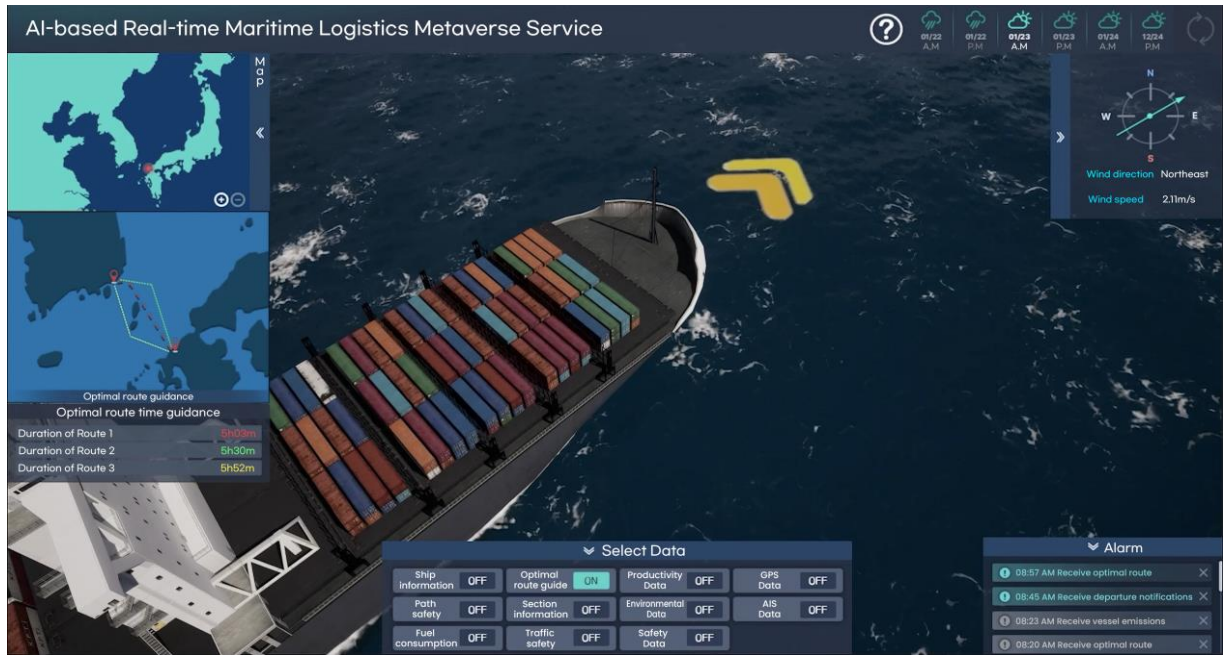

**Figure 4.** Example of ETA prediction service in PLMF.

### 3.1.2. Monitoring and optimization of port operations

The optimization of port operations refers to the overall optimization of the entire port process, from the time a vessel arrives through berthing operations, yard management, and finally, the transfer of goods outside the port via the gate. Port optimization is performed for various purposes, such as equipment allocation, scheduling, and minimizing rehandling, and ranges from optimizing specific tasks to optimizing the entire process. The PLMF metaverse utilizes IoT sensor data and TOS information to monitor various tasks, measure the load and congestion caused by delays, and proposes optimization strategies to reduce these inefficiencies.

Figure 5 illustrates an example screen within the metaverse that monitors berthing operations and presents optimized strategies for enhancing port operation efficiency. The upper section presents the berthing and unloader operation ratios, providing insights into the current status of the berthing operations, task plans, performance, and progress. In addition, relevant data can be selected to review further operational details. Port operation optimization requires the consideration of various factors; therefore, the system is configured based on optimization strategies derived from simulations.

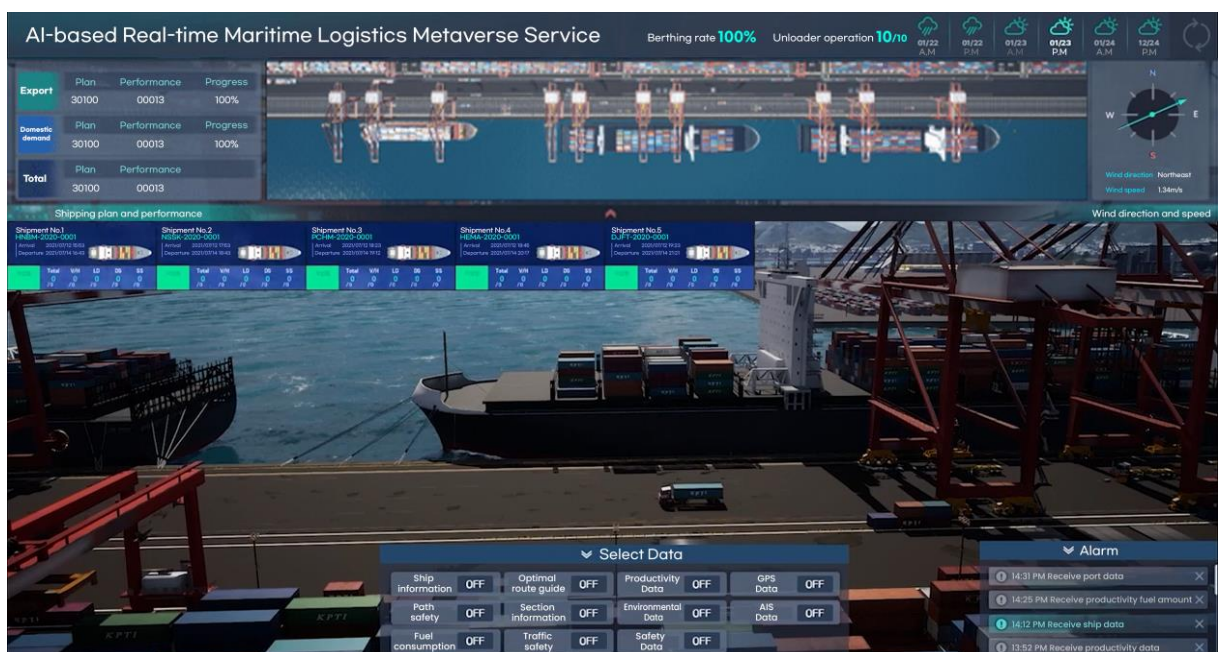

**Figure 5.** Example of a port operation optimization service in PLMF.

## 3.2. PLMF service scenarios from an environmental perspective

### 3.2.1. Monitoring and prediction of fuel consumption

Greenhouse gases emitted by ships are among the main causes of marine pollution in seaport supply chains. The International Maritime Organization (IMO) has developed a roadmap for reducing greenhouse gas emissions from ships and has proposed several policies. Technical measures such as the energy efficiency existing ship index and operational measures such as carbon intensity indicator (CII) regulations have been introduced (Bayraktar & Yuksel, 2023). In particular, the CII strengthens environmental regulations by calculating the efficiency of the ship based on its actual annual fuel consumption and operating distance and subsequently assigning a rating. Therefore, the PLMF metaverse provides a service to monitor the fuel consumption information of operating ships. Additionally, we offer an artificial intelligence service that predicts future fuel consumption using the collected fuel consumption data, anticipated routes, and meteorological information for sailing areas. In this process, we used specialized time-series prediction models, such as correlation recurrent units (CRU), which can account for the correlation and autocorrelation of time-series input data, to ensure accurate predictions (Sim et al., 2023).

Figure 6 shows an example of a service screen within a metaverse that visualizes the real-time fuel consumption of ships and provides predictive outcomes. The center of the figure displays the predicted emission results for the ship. The top-left corner provides the location and route of the ship, whereas the bottom section provides information on the fuel consumption. Similar to other service screens, the top-right corner provides weather information, and the bottom section provides notifications of the necessary information. Based on this, key users of the service, such as navigators and shipping companies, can access the decision-making support information they need to determine routes and fuel consumption to meet the CII ratings.

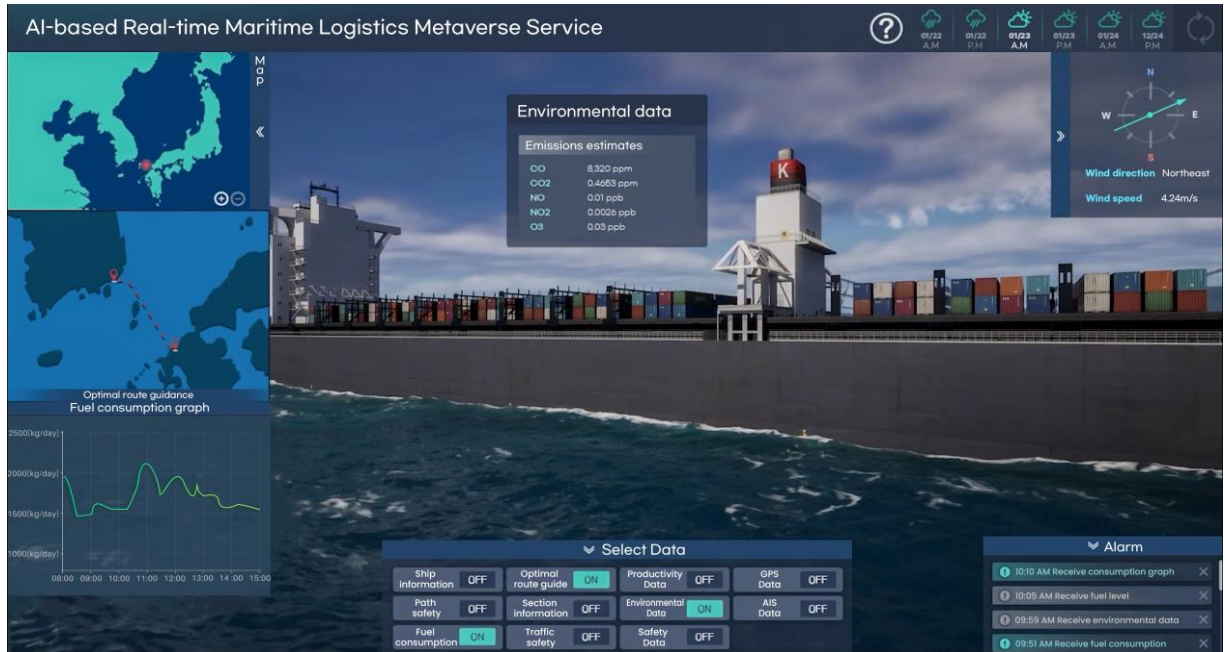

**Figure 6.** Example of a ship fuel consumption monitoring service in PLMF.

### 3.2.2. Monitoring and prediction of port equipment emissions

The major sources of air pollution in ports include ships, cargo-handling equipment, and transfer equipment. Although several studies have focused on estimating and analyzing air pollutant emissions from ships, there is a limited research on monitoring and predicting pollutant emissions from various port equipment. However, with the identification of several port equipment with outdated diesel engines as major sources of pollutants such as CO and NOx, the monitoring, prediction, and evaluation of air pollutant emissions from each piece of equipment has become an important issue (Lee et al., 2023). To address this issue, PLMF metaverse provides a service that monitors air pollutant emissions from port equipment by utilizing IoT sensor data. Additionally, it can predict the dispersion paths of pollutants by considering the contribution of each piece of equipment to the pollution and the spread of air pollution. Furthermore, the system accurately predicts pollution levels in ports by employing a deep collaborative learning model, which not only learns the emission levels of the equipment but also incorporates ship location information near berths and meteorological data from the port in a spatiotemporal context (Sim et al., 2022). Figure 7 illustrates an example of a service screen within the metaverse that provides monitoring and emission estimates for each piece of port equipment. By monitoring the major sources contributing to port pollution and controlling the concentration of pollutants, this service can contribute to improving the air quality in port cities.

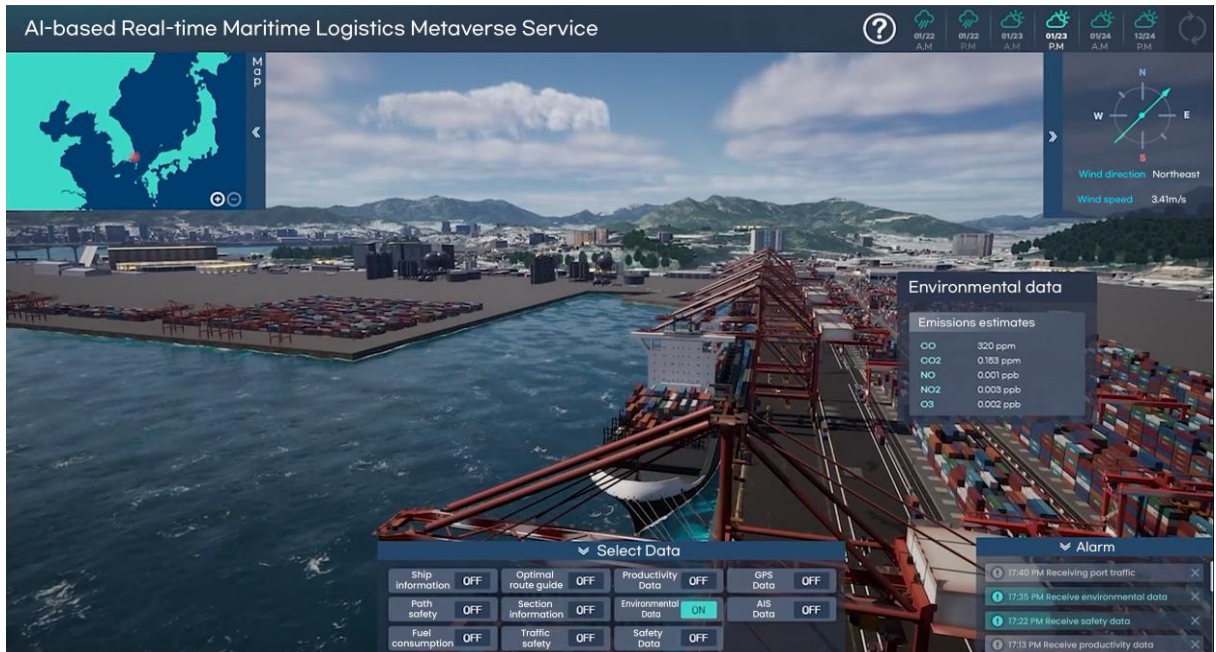
**Figure 7.** Example of hazardous areas detection service in PLMF.

### 3.3. PLMF service scenarios from a safety perspective

#### 3.3.1. Risk assessment of the route
The safety level of a ship can vary significantly depending on various factors such as the characteristics of the navigation area, currents, and underwater conditions. Therefore, it is necessary to evaluate the safety of the route used by the ship in real time and communicate relevant information to stakeholders. In the PLMF metaverse, real-time risk assessment information is provided through reinforcement learning-based simulations based on the location information of the ship and the condition of each maritime area. Figure 8 shows an example of a screen in the metaverse displaying the current risk level associated with a ship's navigation route. On the left side of the screen, the current position of the ship is shown on a map, and a safety graph for the route is displayed, allowing users to decide whether to continue navigating that route. Additionally, by utilizing future route information, we can analyze the risk levels for upcoming routes and provide the necessary information to primary users, such as navigators and ship safety managers.

#### 3.3.2. Detection and monitoring of hazardous ship routes
Although evaluating the risk levels of individual ship routes is important, assessing the risk levels of maritime areas is crucial to the port logistics environment. One of the key advantages of the proposed PLMF metaverse is its ability to accurately display the current positions of ships worldwide based on their location information and provide this data to users.

This capability allowed us to estimate the traffic volume of ships in each maritime area within a metaverse accurately. Using this additional information, we estimated the empirical ship domain (ESD) for each maritime area within the metaverse. Furthermore, hazardous zones can be monitored and detected based on the distribution of ESDs for all ships in the same maritime area. This functionality not only allowed us to accurately communicate the current risk status of specific maritime areas to users within the metaverse but also helped identify potential risks for safer navigation. Figure 9 provides an example screen showing how information about hazardous zones is provided to users based on the traffic volume of ships and the ESD information in each maritime area. The center of the figure displays the route risks within the port, port traffic volume, and risk indicators. Additionally, the left side of the screen shows the current position of the ship and traffic volume for each navigational route by time of day, supporting users in making decisions to prevent accidents in advance.

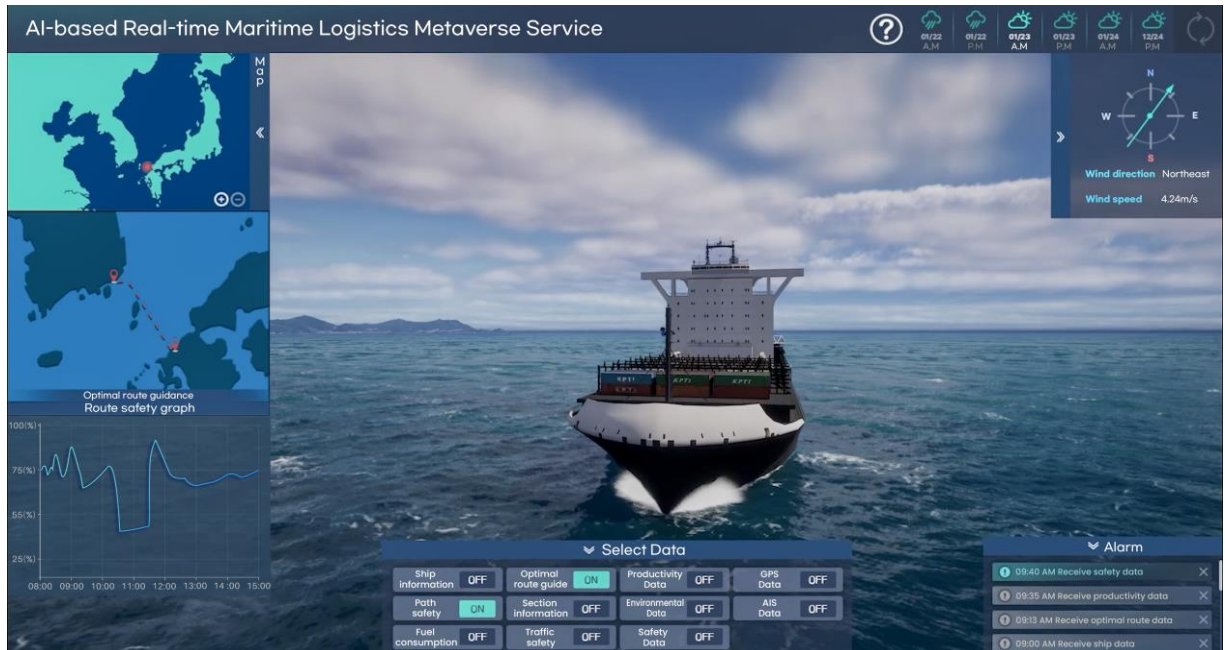
**Figure 8.** Example of ship route risk assessment service in PLMF.

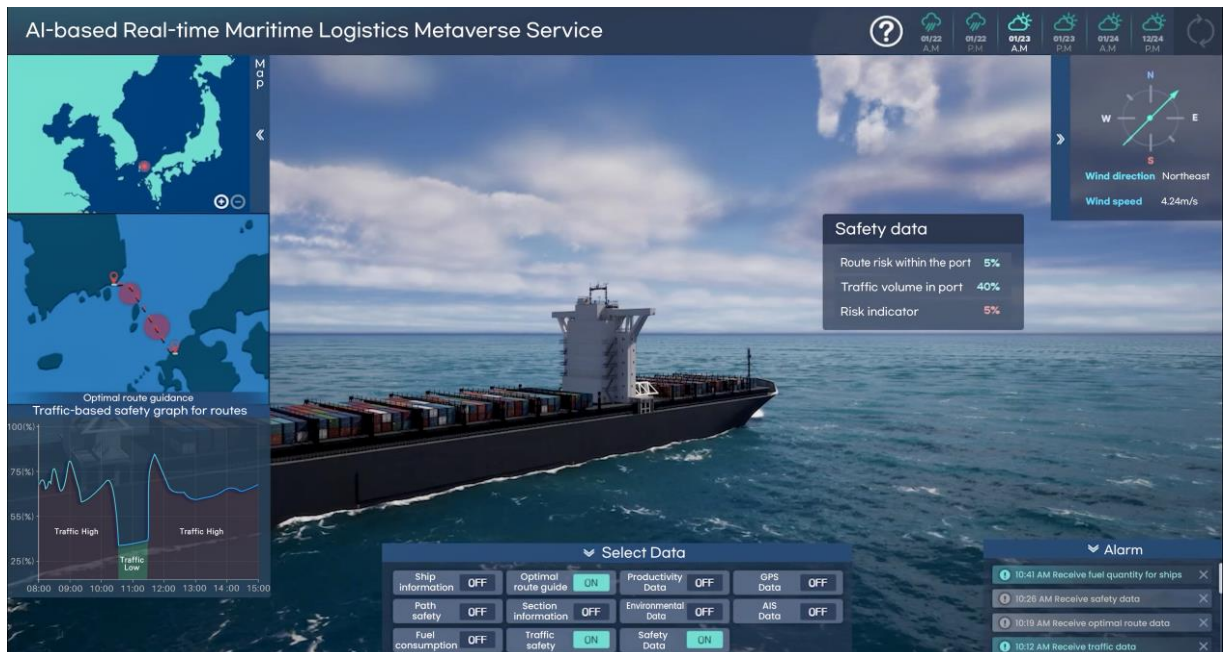
**Figure 9.** Example of hazardous areas detection service in PLMF.

### 3.3.3. Detection and monitoring of accidents between workers and equipment
With the surge in cargo volume, ports are experiencing not only a shortage of shipping capacity but also an increase in unnecessary workloads, which exposes port workers to greater risks and results in an increase in accidents. At Busan Port, institutional measures, such as the Special Port Safety Act, have highlighted the importance of safety. Given the substantial human and material losses caused by port accidents, it is essential to ensure port safety. To address these issues, the PLMF metaverse offers a service that monitors the movement paths of port workers and equipment using GPS information. Additionally, by accurately predicting future paths and locations using the neural ordinary differential equations (neural ODE) methodology, the system can assess potentially hazardous situations and issue warning alerts.

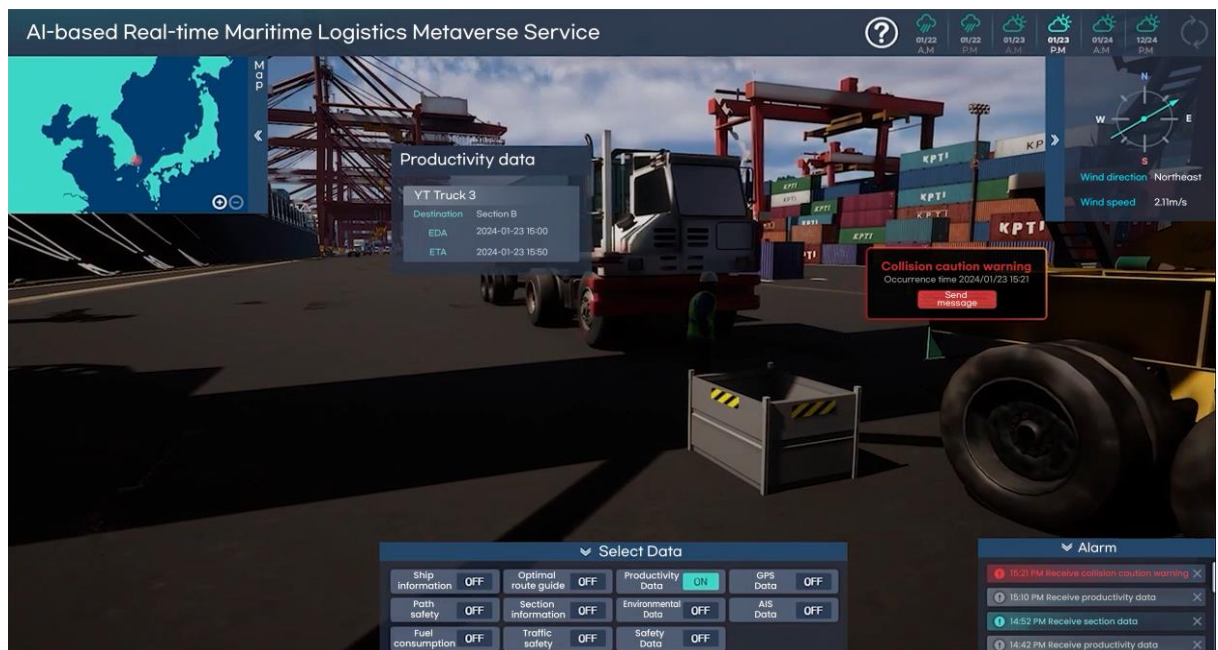

**Figure 10.** Example of accident detection service in port logistics metaverse.

Figure 10 illustrates an example of a service screen within the metaverse that provides real-time alerts about dangerous situations in which a YT might collide with workers. By utilizing information about the route and timing of the YT as well as data regarding the worker, the system can predict potential future accidents and issue collision warnings to workers and truck drivers via alerts. This functionality not only helps users recognize and prevent accidents, but also contributes to mitigating financial losses and supply chain disruptions caused by port accidents.

## 4. Case study: Enhancements in productivity and environmental impact at Busan Port through PLMF

In this section, a case study is conducted using the Busan Port to validate the impact of the PLMF on actual port logistics. Generally, before departing from a port, a shipping company communicates the vessel's operational schedule with the relevant port. Upon receiving this schedule, the port establishes its own operational plans (such as berth planning, equipment planning, and work scheduling) to handle cargo. Once the berth plan is established, it is communicated back to the shipping company, which then determines the ETA based on this plan, and navigates to the port to arrive at the given ETA. However, 70% of the vessels worldwide are unable to arrive within the given ETA owing to external weather factors, environmental conditions, or other issues. Consequently, the vessel may not dock at port within the specified ETA, resulting in unnecessary waiting times. This issue negatively affects port productivity and the environment.

We simulated and analyzed the process within the PLMF, as illustrated in Figure 7, which includes 1) detecting the RTA, 2) estimating a new ETA when an RTA is detected, and 3) revising the berth plan based on the estimated ETA. For this analysis, we used AIS data collected from vessels arriving and departing from Busan Port between January 2021 and December 2021, along with IoT, PCS, and POS data collected from Busan Port during the same period.

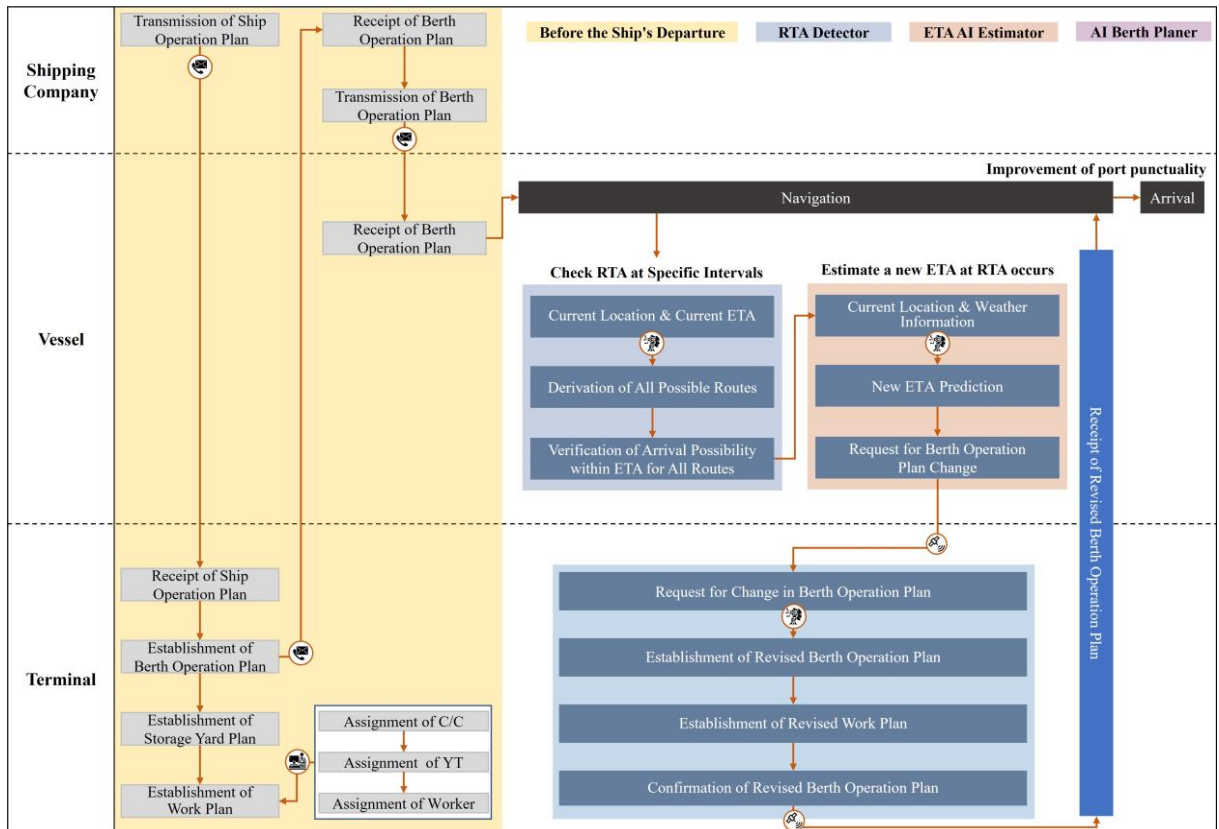

**Figure 7.** Example of hazardous areas detection service in maritime logistics metaverse.

## 4.1. Prediction Model Framework

In this case study, we developed a multichannel ConvLSTM model to determine the ETA, as shown in Figure 8. The input data for the multichannel ConvLSTM are generated as follows. 1) The ship is represented as a grid matrix of a specific size based on its current position. 2) The ship movement path and weather information along this path were combined into a multichannel input matrix for a given time point. 3) The data for the desired time points (t) were aggregated to create a single-input dataset (*X*). In addition, the actual arrival time at the current position of the vessel was used as the output data (*Y*) for training.

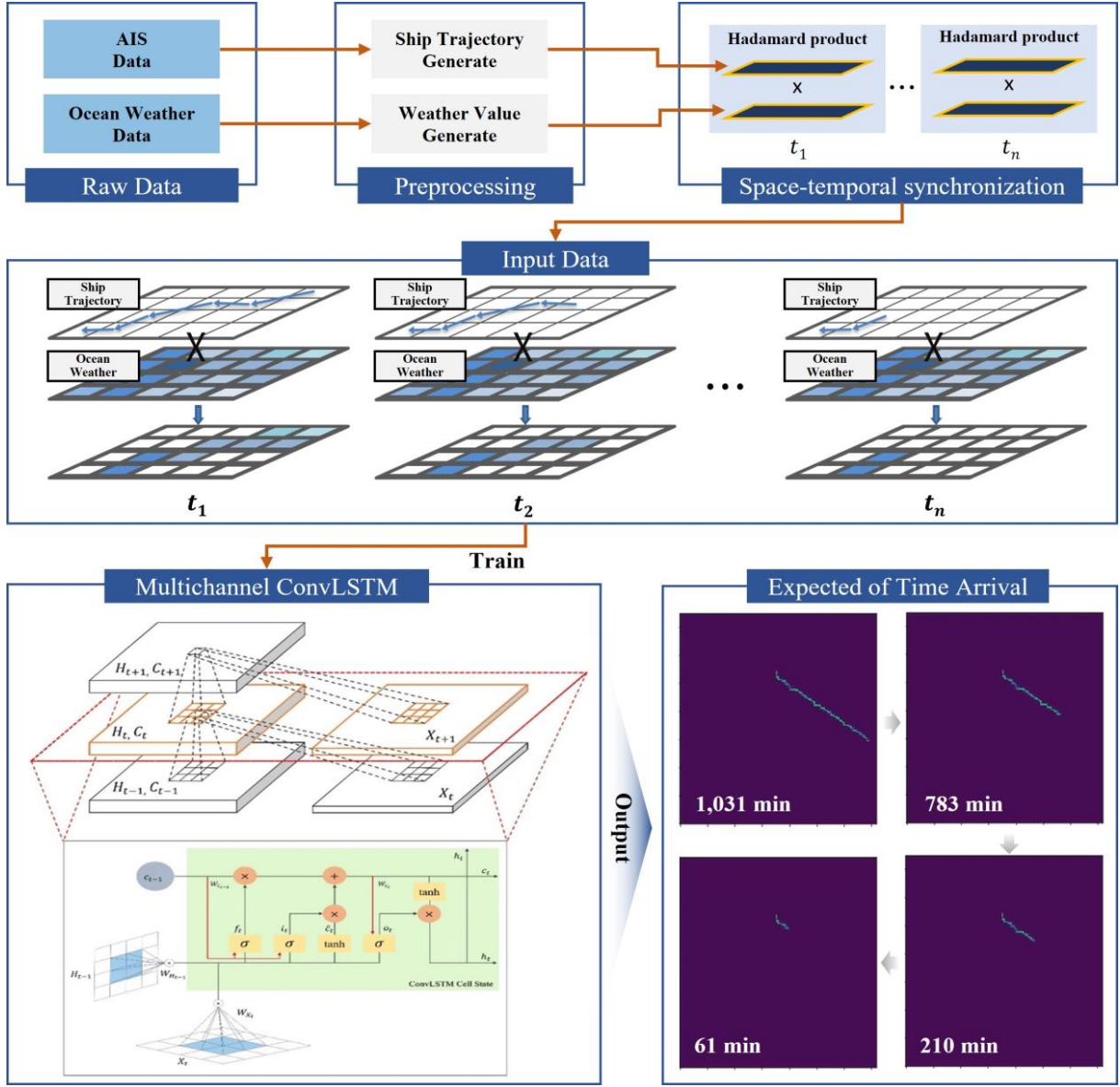

Figure 8. Framework of the Multi-channel ConvLSTM for Detection RTA and Prediction ETA

To compare the performance of the models, we implemented and compared the original ConvLSTM, convolutional neural networks (CNN) + LSTM, and CNN models. We then compared the prediction performance of the weather-informed multichannel ConvLSTM with these models. The Root Mean Square Error (RMSE) and Mean Absolute Percentage Error (MAPE) were used as the comparison metrics.

Table 7 presents the comparison results of the model performances with respect to the accuracy of the estimated arrival times. Overall, comparing the results when the AI models were applied within the PLMF versus when they were not applied, the results showed improvements in all cases. Additionally, the multichannel ConvLSTM model, which considers weather effects, demonstrated an approximately 9.5-fold improvement in the prediction error compared to when no predictive model was applied. Furthermore, it showed performance improvements ranging from a minimum of 20% to up to 3.5 times better those of the models used in previous research.

Table 7. Prediction errors of ETA models compared to actual observations when applying the ETA prediction model within the PLMF.

| Model | RMSE (min) | MAPE (%) |
|---|---|---|
| **Multi-Channel ConvLSTM*** | **27.05** | **3.71%** |
| **Multi-Channel ConvLSTM* (>100km)** | **41.97** | **6.12%** |
| ConvLSTM | 35.17 | 5.96% |
| CNN+LSTM | 68.19 | 8.31% |
| CNN | 128.33 | 13.61% |
| Without model application | 295.61 | 28.31% |

### 4.2. Simulation Analysis

To analyze the cost impact of the PLMF, we conducted a simulation analysis using the following procedure: 1) Assume a situation in which ships that are supposed to arrive on time are intentionally delayed (by generating random RTAs). 2) The ETA is predicted using the multichannel ConvLSTM model when an RTA occurs. 3) The predicted results are reflected in the berth schedule.

We compare the productivity (berth throughput) when the port operates according to the updated berth plan with that when it operates based on the original berth plan without ETA predictions. The comparison focuses on two scenarios: one in which the ETA is immediately predicted and the berth plan is revised accordingly, and the other in which the port operates based on the original berth plan without updating the ETA predictions. Additionally, simulations were conducted by varying the RTA occurrence rates, incrementing them from 5% to 30% at 5% intervals.

Table 8. Simulation results based on RTA occurrence rates (results obtained with the ETA prediction model applied vs. results obtained without the ETA prediction model).

| Rate of RTA Occurrences | Throughput per Hour (Processing Time per 1 VAN) | |
|---|---|---|
| | Without Applying the Predicting ETA Model | With Applying the Predicting ETA Model |
| 5% | 27.77van (129.6 sec) | 27.96van (128.7 sec) |
| 10% | 27.56van (130.6 sec) | 27.92van (128.5 sec) |
| 15% | 27.38van (131.4 sec) | 27.89van (129.0 sec) |
| 20% | 27.20van (132.3 sec) | 27.83van (129.3 sec) |
| 25% | 27.00van (133.3 sec) | 27.76van (129.7 sec) |
| 30% | 26.82van (134.2 sec) | 27.67van (130.1 sec) |

Table 8 lists the simulation results. The values in the table represent the number of containers handled by a quay crane per hour. As the RTA occurrence rate increased, the throughput decreased in both cases (with and without the ETA prediction model). At an RTA occurrence rate of 30%, applying the ETA prediction model allows one quay crane to handle 0.85 more containers. Table 9 compares the daily throughput considering the quay crane availability at "A Terminal" in Busan Port, where the case study was conducted. Analysis of the POS data for "A Terminal" revealed that 15 cranes are available, with an average of 13.5 cranes in operation. Considering this, applying the ETA prediction model results in handling approximately 265–285 more containers per day compared with not applying the model. Over the course of one year, this translates to handling an additional 96,72 van to 104,025 vans. Considering the average value of about $70 per container (1 van) at "A Terminal", this results in an additional benefit of up to $7,290,080. Based on these findings, applying the PLMF to port logistics operations can contribute to increased productivity and profit generation.

Table 9. Analysis of additional revenue based on the simulation results.

| Num of Quay Crane | Comparison of Quay Side Productivity (30% RTA Occurrence Rate) | | | | |
|---|---|---|---|---|---|
| | When processing 26.82van (a) | When processing 27.67van (b) | 1Day Difference (b-a) | 1Year Difference | Additional Revenue (70$ per 1van) |
| 1 | 643.7 | 664.1 | 20.4 | 7446 | 521220 |
| 2 | 1287.4 | 1328.2 | 40.8 | 14892 | 1042440 |
| 3 | 1931.0 | 1992.2 | 61.2 | 22338 | 1563660 |
| 4 | 2574.7 | 2656.3 | 81.6 | 29784 | 2084880 |
| 5 | 3218.4 | 3320.4 | 102.0 | 37230 | 2606100 |
| 6 | 3862.1 | 3984.5 | 122.4 | 44676 | 3127320 |
| 7 | 4505.8 | 4648.6 | 142.8 | 52122 | 3648540 |
| 8 | 5149.4 | 5312.6 | 163.2 | 59568 | 4169760 |
| 9 | 5793.1 | 5976.7 | 183.6 | 67014 | 4690980 |
| 10 | 6436.8 | 6640.8 | 204.0 | 74460 | 5212200 |
| 11 | 7080.5 | 7304.9 | 224.4 | 81906 | 5733420 |
| 12 | 7724.2 | 7669.0 | 244.8 | 89352 | 6254640 |
| 13 | **8367.8** | **8633.0** | **265.2** | **96798** | **6775860** |
| 14 | **9011.5** | **9297.1** | **285.6** | **104244** | **7297080** |
| 15 | 9655.2 | 9961.2 | 306.0 | 111690 | 7818300 |

Table 10. Comparison of ship punctuality based on the simulation results.

| Ship Punctuality (Estimated Arrival Time - Actual Arrival Time) | Without Applying the Predicting ETA Model | With Applying the Predicting ETA Model |
|---|---|---|
| Mean | 121.9 min | 25.4min |
| Median | 45.0 min | 12 min |
| Standard Deviation | 265.1 min | 147.3 min |

We further examined whether applying the ETA prediction model within the PLMF would improve ship punctuality. According to the results listed in Table 10, when the ETA prediction model was not applied, there was an average discrepancy of approximately 2 h between the actual arrival time and ETA. However, when the ETA prediction model was applied, this discrepancy was reduced to approximately 25 min. This demonstrates that the PLMF improves the punctuality of actual ship arrivals. If a ship fails to maintain punctuality, it cannot begin operation at the originally scheduled time. Consequently, the ship must wait at anchorage before berthing. Several studies have reported that waiting times at anchorages contribute significantly to pollution in port cities. PLMF, by significantly improving ship punctuality, also helps reduce environmental pollution in ports. Figure 9 shows the waiting times of ships near the port before and after applying the ETA prediction model. The data illustrated in Figure 9 reveals a significant reduction in waiting times of ships near ports, indicating that the PLMF can also contribute to mitigating environmental issues arising from port logistics.

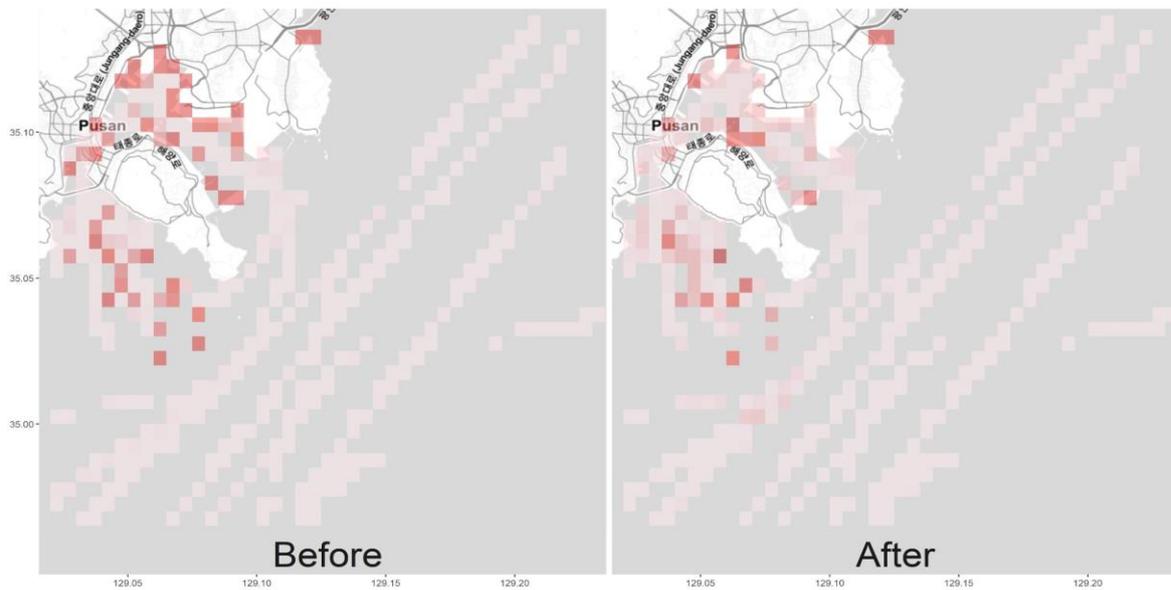

Figure 9. Visualization of differences in ship waiting times before and after applying the ETA prediction model within the PLMF.

## 5. Conclusions

This study proposes an AI-based PLMF designed to address productivity, environmental, and safety issues in port logistics. The proposed PLMF utilizes data collected from various systems, including AIS, port IoT devices, PCS, POS, and PAQMS to monitor port logistics processes occurring in the sea, berth, and port areas. This framework is structured to contribute significantly to data sharing, service connectivity, and interactions among various entities and stakeholders in port logistics by considering the data generated from all systems used in the port logistics process.

Another notable feature of the PLMF is the integration of various AI models to enhance the accuracy of port logistics decision making. Several AI models have been applied within the PLMF, including CNN, ConvLSTM, CNN+LSTM, neural ODE, CRU, heuristic approaches, and reinforcement learning. These models provide metaverse content to address four productivity issues (detecting required time of arrival; predicting expected time of arrival; dynamic berth planning; and dynamic port operation planning), three environmental issues (prediction and monitoring of ship emissions, near-port pollution, and port emissions), and three safety issues (ship route risk assessment; prediction and monitoring of ship collisions; and detection and monitoring of port safety).

We conducted a case study using historical data from Busan Port to evaluate the effectiveness of the PLMF. Using the PLMF, we predicted the arrival times of ships and simulated the optimization of port operations based on these predictions. We observed that the framework could generate approximately 7.3 million dollars in additional direct revenue annually and improve ship punctuality by 79%, resulting in environmental benefits for the port. These findings demonstrate that the PLMF not only provides a platform for collaboration among various stakeholders in port logistics but also significantly enhances the accuracy and sustainability of decision-making through AI-based simulations.

The PLMF is expected to enhance the value of the logistics industry by monitoring all port logistics processes in real-time and improving productivity through accurate AI-based decision making. Additionally, it aims to address environmental and safety issues occurring at ports, thereby contributing to overall improvements in the port logistics sector.

**Acknowledgments**

This work was supported by the National Research Foundation of Korea (NRF) grant funded by the Korean government (MSIT) (No.RS-2023-00218913) and by MSIT, Korea, under the Grand Information Technology Research Center support program (IITP-202-2020-0-01791)